\documentclass[final]{article}

\usepackage{graphicx}
\usepackage{hyperref}

\begin{document}

\title{Progress in Top Quark Physics\footnote{Submitted to
Proceedings of PANIC05, Particles and Nuclei International Conference, 
Santa Fe, New Mexico, October 24-28, 2005.}}

\author{Evelyn J. Thomson \\
Department of Physics and Astronomy, University of Pennsylvania.\\
}
\date{}

\maketitle

\begin{abstract}

Experimental measurements of the properties of the top quark
have improved and will continue to improve significantly, 
with the excellent operation of the CDF and D0 experiments and 
the Tevatron $p\bar{p}$\ collider at the Fermi National 
Accelerator Laboratory.  
All of the final state experimental signatures from top quark production
and decay are being analysed to test if this most massive quark is sensitive
to new physics beyond the standard model.  So far, observations are consistent
with the standard model. New techniques have dramatically improved 
the precision of the top quark mass measurement to 1.7\%  and set the stage
for a sub-1\% measurement by 2008.  This improved knowledge of the 
top quark mass sharpens the standard model prediction for the mass 
of the undiscovered Higgs boson, with implications for 
Higgs studies at the future LHC and ILC.

\end{abstract}


\section*{Introduction}

The top quark~\cite{pdgtopquark,jacoreview} was discovered in 1995 by CDF and D0 
at the $\sqrt{s}=1.8$~TeV $p\bar{p}$\ Tevatron collider~\cite{cdfdiscovery,d0discovery}. The top quark is the most
massive fundamental particle in the standard model of particle physics, 
with a mass approximately twice that of the $W$ and $Z$ bosons, 
the carriers of the electroweak force, and thirty-five
times that of the next most massive fermion, the $b$ quark.  The standard model neither
predicts nor explains the observed mass hierarchy.  The large mass of the top quark
implies a unique large coupling to the elusive standard model Higgs boson.  
Precision studies of top quark properties
have the potential to reveal effects from theories beyond the standard model. 
Until recently, this potential has been limited by low statistics, 
with only a few dozen candidates in the 100~pb$^{-1}$\ collected through 1995.  

This general review covers representative recent results on top quark physics from the CDF and D0 
experiments with 350~pb$^{-1}$.  Due to limited space, it cannot cover all
of the many results~\cite{CDFD0TOP}. 

\section*{Top Quark Pair Production}

Top quarks are produced in pairs via the strong interaction processes
$q\bar{q} \to t\bar{t}$\ and $gg \to t\bar{t}$.  
The prediction from QCD at next-to-leading order in $\alpha_{s}$\ 
for the top quark pair production cross section~\cite{ttbarMangano,ttbarKidonakis} is 
$6.7 \pm^{0.7}_{0.9}$~pb for $m_{\mathrm{top}}=175$~GeV/$c^{2}$.
Due to the increase in $\sqrt{s}$\ to 1.96~TeV, this is 30\% higher than at 1.8~TeV. 
The theoretical uncertainty includes scale and parton distribution function variations.  
Note that the production cross section depends strongly on the top quark mass, 
increasing to $7.8\pm^{0.9}_{1.0}$~pb for $m_{\mathrm{top}}=170$~GeV/$c^{2}$.

In order to produce sufficient numbers of top quarks for precision studies, 
this tiny top quark pair production cross section necessitates the operation 
of the highest energy hadron collider in the world 
at the highest luminosity in the world.  
Current results from the CDF and D0 experiments at the Tevatron collider
are based on detailed analysis of 
approximately 350~pb$^{-1}$\ of $p\bar{p}$\ collisions at $\sqrt{s}$=1.96~TeV 
collected between 2002 and 2004.  
In 2005, the Tevatron reached peak instantaneous luminosities over 
150$\times 10^{30}$cm$^{-2}$s$^{-1}$, 
and the CDF and D0 experiments each accumulated over 700~pb$^{-1}$.
Instantaneous or integrated, the operation of the upgraded Tevatron 
is now over seven times better than for the 1994-1995 top quark discovery run!  
The future is even brighter: electron cooling~\cite{electroncooling} has begun 
to be used to reduce the size of the anti-proton beams prior to injection 
into the main Tevatron collider.  This is the first application of electron cooling 
at relativistic beam energies, and also the first time the technique has been used 
concurrently with stochastic cooling.  The ultimate potential of electron cooling is 
to increase the instantaneous luminosity by 50-100\%, and to increase the expected 
integrated luminosity per experiment from 4~fb$^{-1}$\ to as much 
as 8~fb$^{-1}$\ by the end of 2009.  

Having gone to all this effort to produce top quarks, their existence is truly fleeting.
With a lifetime of the order of 10$^{-25}$~s, there is not even enough time for 
the top quark to hadronise into mesons or baryons, unlike any other quark.   
Therefore, the spin of the top quark should be 
preserved in the angular distribution of the top quark decay products.
In the standard model, the top quark decays 
via the electroweak interaction to a $W$ boson and a $b$ quark with a 
branching fraction of 99.8\%.  The total decay width is 1.5~GeV for 
$m_{\mathrm{top}}$=175~GeV/$c^{2}$. Note that the top quark is massive enough 
that it could also decay to new exotic particles, 
such as a charged Higgs boson, that have not been excluded yet by direct searches.

There are three experimental signatures from 
standard model top quark pair production and decay, 
$t\bar{t} \to W^{+}bW^{-}\bar{b}$, that are characterized by the number and type 
of charged leptons from the decay of the $W^{+}$\ and $W^{-}$\ bosons:

\begin{itemize}

\item \textbf{Dilepton (branching fraction 10.3\%):} 
Both $W$ bosons decay to a lepton and a neutrino.
The experimental signature is two isolated leptons with opposite electric charge, 
significant missing transverse energy from two undetected neutrinos, 
and at least two jets with large transverse energy originating from the 
two $b$ quarks. 

\item \textbf{Lepton + Jets (branching fraction 43.5\%):} 
One $W$ boson decays to a lepton and a neutrino,
the other $W$ boson decays to a quark and an anti-quark.  
The experimental signature is one isolated lepton, 
significant missing transverse energy from the undetected neutrino, 
and at least four jets with large transverse energy, 
with two of the jets originating from $b$ quarks.

\item \textbf{All hadronic (Branching fraction 46.2\%):} 
Both $W$ bosons decay to $q\bar{q'}$.  
The experimental signature is at least six jets with large transverse energy, 
with two of the jets originating from $b$ quarks.


\end{itemize}

In practice, the challenge of filtering off for later analysis 
less than 100~Hz of the 1.7~MHz $p\bar{p}$ bunch collision rate
makes collecting a sample of top quarks a bit like trying to pan for 
gold under Niagara Falls! The huge $p\bar{p}$\ cross section
of 60~mb and the high instantaneous luminosity mean that on average there is at least one 
$p\bar{p}$\ interaction per bunch collision that
leaves measurable energy in the detectors.  However, only one in ten billion
$p\bar{p}$\ interactions produces a brace of top quarks.  Fortunately, an electron or muon with 
high transverse energy provides a highly efficient way to trigger on 
the dilepton and lepton+jets channels.  Although the background from 
QCD multi-jet production is immense, the large number and high transverse energy
of the jets in the all-hadronic channel also allows a trigger with high efficiency and
low enough rate for the available bandwidth.

\subsection*{Dilepton}

D0's basic selection requires two isolated, identified leptons (electrons or muons)
with $p_{T}>15$ GeV/$c$, and at least two jets with $p_{T}>20$ GeV/$c$ 
reconstructed by a cone algorithm with radius $\Delta \mathcal{R}$=0.5.
The pseudo-rapidity ranges are $|\eta|<2.5$ for jets, $|\eta|<2.0$ for muons, 
and $|\eta|<2.5$\ excluding the range $1.1 < |\eta| < 1.5$\ for electrons.

While the background process $Z/\gamma^{\star} \to \ell^{+}\ell^{-}$ with associated jets
has a much larger production cross section than $t\bar{t}$,  it can be reduced
in the $e^{+}e^{-}$\ and $\mu^{+}\mu^{-}$\ channels 
by exploiting both the peak in dilepton invariant mass at the $Z$ boson mass,
and the smaller missing transverse energy due only to the finite resolution on the 
measurement of jet energies.  
In the case of $Z/\gamma^{\star} \to \tau^{+} \tau^{-}$ with associated jets, 
with subsequent $\tau \to e\nu_{e}\nu_{\tau}$ 
or $\tau \to \mu\nu_{\mu}\nu_{\tau}$ decay, 
instead the lower $p_{T}$ of the charged leptons and the jets 
allows discrimination.  Other sources of background include diboson production, 
and fakes, which are mainly from $W$\ boson production with associated jets where 
a hadronic jet is mis-identified as a lepton.  

\begin{figure}
\begin{minipage}{0.48\textwidth}
\includegraphics[width=.99\textwidth,bb= 0 0 600 600]{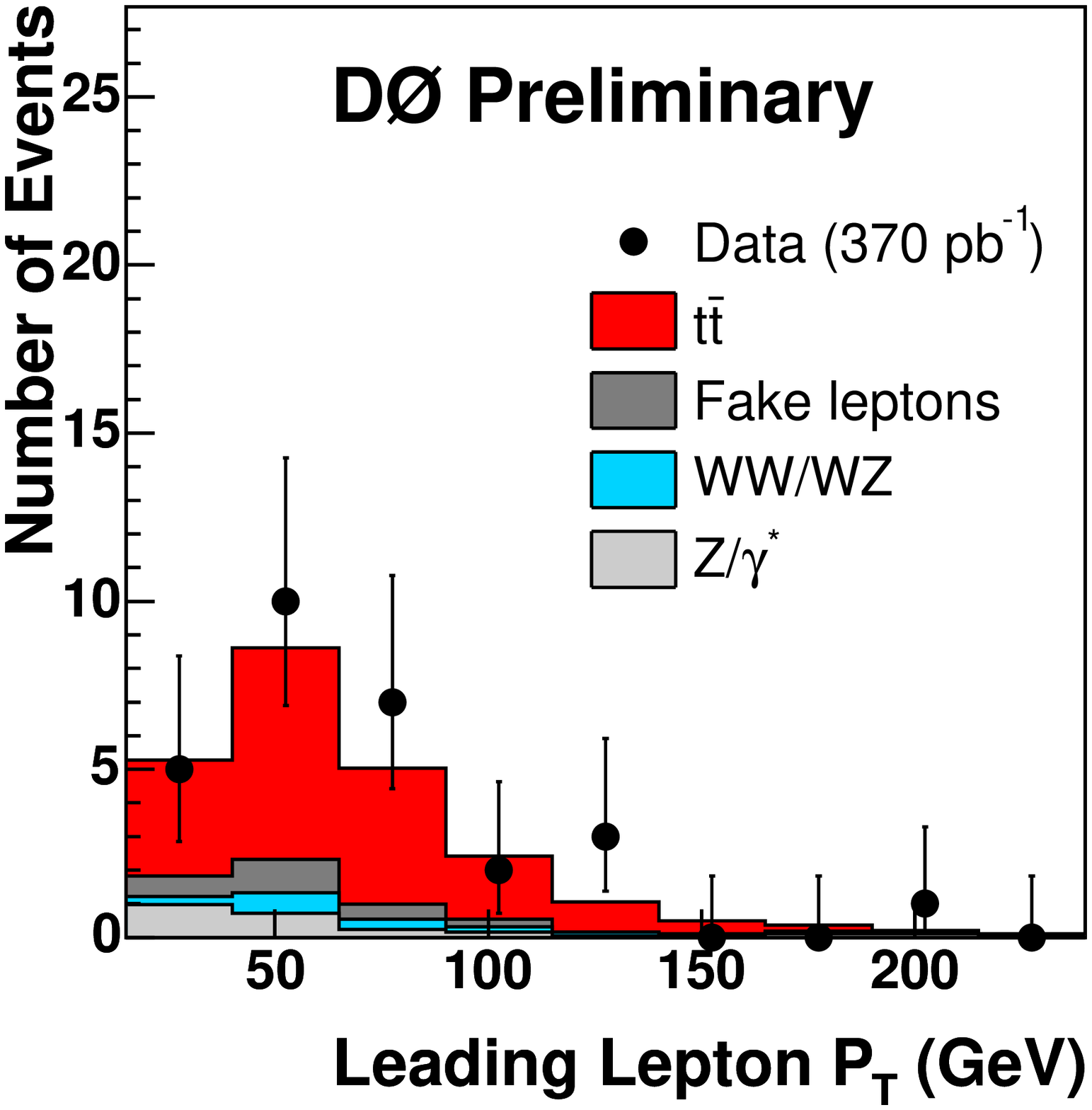}
  \caption{\label{fig:d0_leptonpt}The $p_{T}$\ distribution 
of the highest $p_{T}$\ electron or muon in the D0 dilepton channel.  
The contribution from $t\bar{t}$\ is normalized to 7~pb.}
\end{minipage}
\hspace{0.03\textwidth}
\begin{minipage}{0.48\textwidth}
\includegraphics[width=0.99\textwidth]{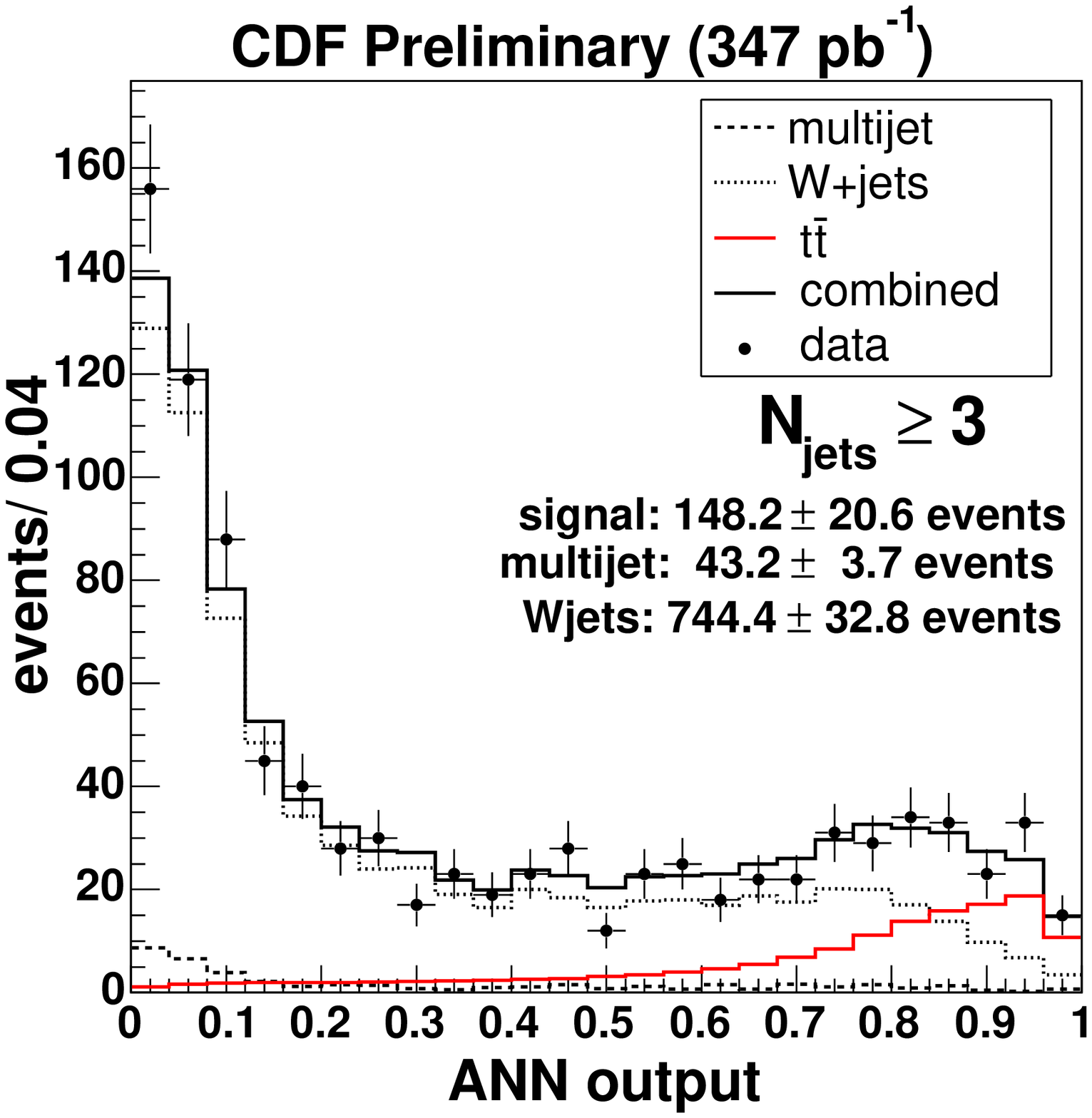}
  \caption{\label{fig:cdf_ann}The artificial neural network output
in the CDF lepton+jets channel.  The best fit 
for the contribution from $t\bar{t}$\ is shown.}
\end{minipage}
\end{figure}

After all selection requirements, 
the efficiency times branching fraction is about 0.7\% for $t\bar{t}$.  
D0 observes 28 events in 370~pb$^{-1}$,
with an estimated background of 6.8$\pm$2.2 events. Figure~\ref{fig:d0_leptonpt} 
compares the observed $p_{T}$\ of the highest $p_{T}$\ lepton with the expectation 
from the standard model.
D0 measures $\sigma_{t\bar{t}} = 8.6\pm^{2.3}_{2.0}\pm^{1.2}_{1.0}\pm0.6$~pb~\cite{d0dileptonxs},
where the first uncertainty is statistical, the second due to systematic uncertainties 
on the signal efficiency and the background estimate, 
and the third from the uncertainty on the integrated luminosity.
With 360~pb$^{-1}$, CDF measures $\sigma_{t\bar{t}} = 10.1\pm2.2\pm1.3\pm0.6$~pb~\cite{cdfdileptonxs}.

CDF searches for the $e\tau$\ and $\mu\tau$\ final states, where the $\tau$\
decays hadronically.  Although the selection efficiency times branching fraction is only 0.08\%,
this final state could be enhanced by effects from new physics, \emph{e.g.}  
if the top quark decays via a charged Higgs boson, $t \to H^{+}b$, 
with $H^{+} \to \tau^{+}\nu_{\tau}$.  In 195~pb$^{-1}$, CDF observes 2 events on 
a background of 1.3$\pm$0.1 events.  CDF sets a 95\% confidence level limit 
that the $t \to \tau\nu_{\tau}q$\ branching fraction
is no higher than 5.2 times the standard model branching fraction~\cite{cdfelltau}.

\subsection*{Lepton + Jets}

CDF's basic selection requires exactly one isolated identified lepton (electron or muon)
with $p_{T}>$20~GeV/$c$, missing transverse energy above 20~GeV, and at least
three jets with $E_{T}>$15~GeV reconstructed by a cone algorithm 
with radius $\Delta \mathcal{R}$=0.4.  The pseudo-rapidity ranges are $|\eta|<2.0$ for jets
and $|\eta|<1.1$ for leptons.  The efficiency times branching fraction 
is about 7\% for $t\bar{t}$.  

The decay products of the massive top quark are more energetic and central than
those from the main background of $W$ boson production with associated jets.  CDF
combines the discriminating power of several kinematic and angular event observables 
in an artificial neural network.  Figure~\ref{fig:cdf_ann} shows the neural network
output for 936 observed data events in 347~pb$^{-1}$.  
The $t\bar{t}$\ distribution is modeled by PYTHIA Monte Carlo,
the $W$+jets distribution by ALPGEN $W+3$~parton matrix-element interfaced with HERWIG.
Background from multi-jets, in which a hadronic jet is mis-identified as a lepton,
is determined from data where the lepton is not isolated.
The best fit to the data prefers 148.2$\pm$20.6 events from $t\bar{t}$,
where the uncertainty is statistical only.
CDF measures $\sigma_{t\bar{t}} = 6.3 \pm 0.8 \pm 0.9 \pm 0.4$~pb~\cite{cdfljxs},
where the dominant systematic uncertainty is from the dependence of the $W$+jets
background shape on the Monte Carlo $Q^{2}$\ scale.
With 230~pb$^{-1}$, D0 measures $\sigma_{t\bar{t}} = 6.7\pm^{1.4}_{1.3}\pm^{1.6}_{1.1}\pm0.4$~pb~\cite{d0ljxstopo}.

Due to their long lifetime and large boost, 
the $B$ hadrons resulting from top quark decay 
travel several millimeters from the primary interaction point before decaying into several particles.
While there are two jets originating from $b$ quarks in each $t\bar{t}$\ event,
only a few \% of the $W$+jets background contains any jets from $b$ or $c$ quarks.
Therefore the ability to identify (tag) a jet containing a $b$-quark 
provides a distinctive experimental signature that can be used
to reduce the background from $W$+jets.  The most powerful $b$-tag algorithm 
requires the direct reconstruction of a secondary vertex from
the charged decay products of the $B$ hadron, where this secondary vertex is 
significantly displaced along the direction of the parent jet 
from the primary $p\bar{p}$\ interaction point.  The typical efficiency of this algorithm
is 45\% for a $b$-jet with $E_{T}>40$ GeV.  The typical false positive (mistag) rate
is 0.5\%.  Note that both the efficiency and the mistag rate depend on the jet $E_{T}$ and $\eta$.

\begin{figure}
\begin{minipage}{0.49\textwidth}
\includegraphics[width=.99\textwidth]{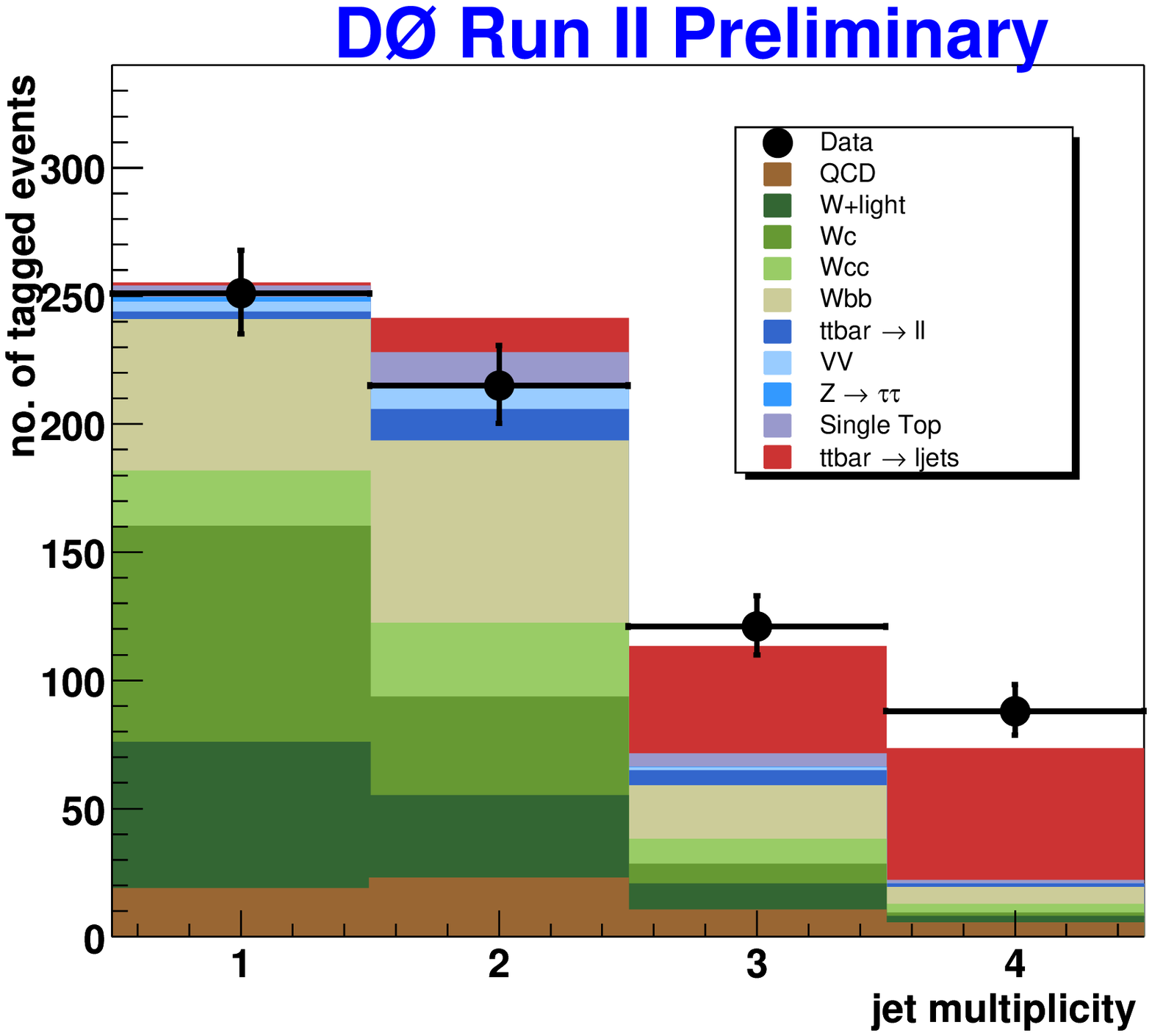}
\end{minipage}
\hspace{0.02\textwidth}
\begin{minipage}{0.49\textwidth}
\includegraphics[width=.99\textwidth]{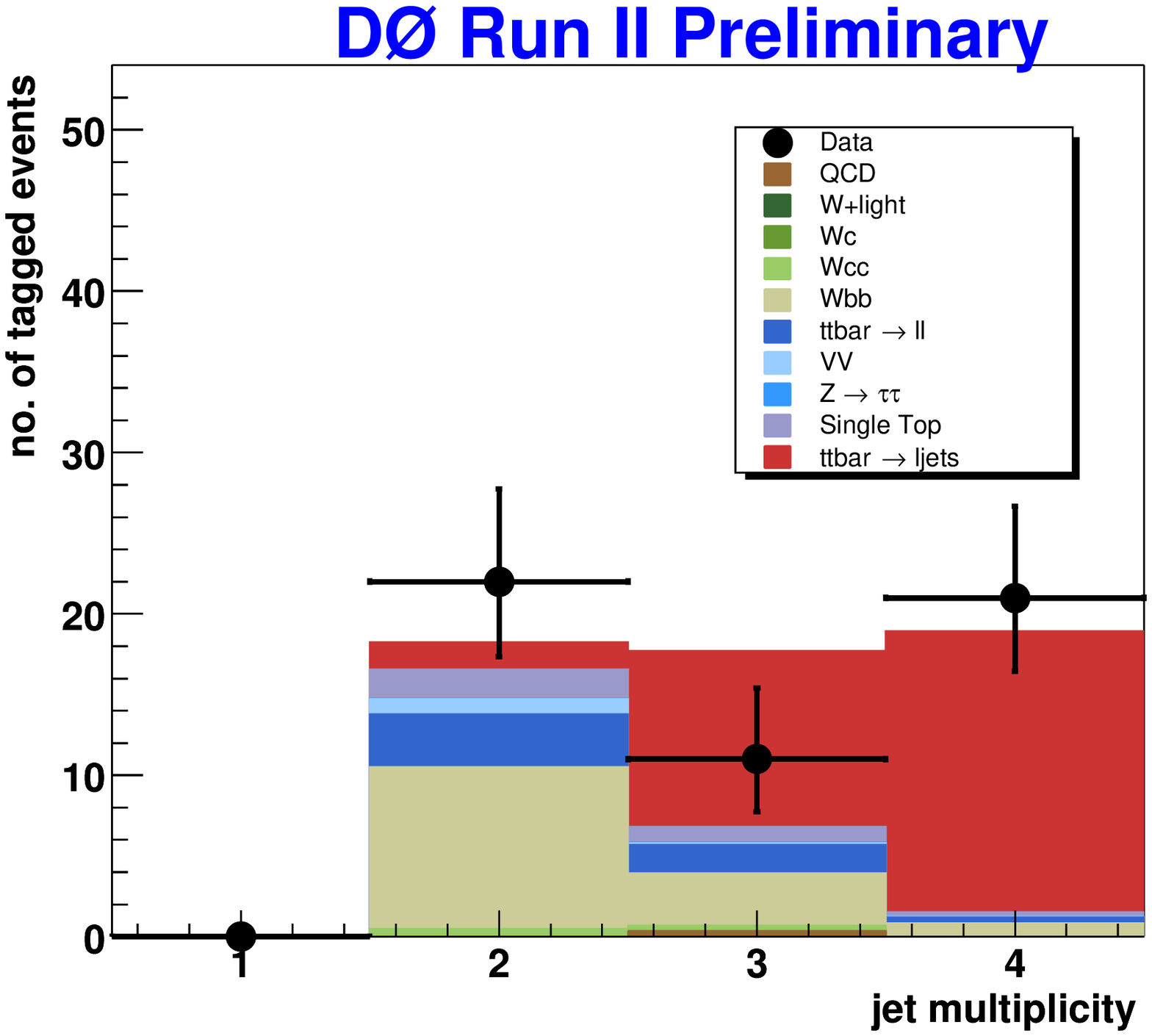}
\end{minipage}
  \caption{\label{fig:d0_btag} The number of predicted and observed
events with a single $b$-tag (left) and two or more $b$-tags (right)
as a function of jet multiplicity in the D0 lepton+jets channel. 
The contribution from $t\bar{t}$\ is normalized to 7~pb.}
\end{figure}

With the requirement of at least one (two) $b$-tags,
the efficiency times branching fraction for $t\bar{t}$\ is about 4\% (1\%). 
In 365~pb$^{-1}$, D0 observes 209 events with exactly one $b$-tag for 
a background estimate of 93$\pm$10 events, and 32 events with at least 
two $b$-tags for a background estimate of 9$\pm$1 events.  
As is evident from Figure~\ref{fig:d0_btag}, which 
compares the observed number of events with exactly one $b$-tag and two or more
$b$-tags with the expectation from the standard model, 
the background estimate contains contributions from many different processes.  
The dominant contributions from $Wb\bar{b}$, $Wc\bar{c}$, and $Wc$ production
with associated jets are based on leading order in $\alpha_{s}$\ Monte Carlo estimates of their 
rates relative to inclusive $W$+jets production.  The absolute normalization is taken
from the number of observed $W$+jets events before $b$-tagging.  This procedure side-steps
the large uncertainty (50\%) on the leading order in $\alpha_s$\ prediction of the 
absolute production rate of $W$+jets. Comparison of the number of observed and 
predicted events in the $W\!+\!1$\ jet and $W\!+\!2$\ jet regions, 
where little contribution from $t\bar{t}$\ is expected, 
provides an important cross-check of the background estimate.
D0 measures $\sigma_{t\bar{t}} = 8.1 \pm 0.9 \pm 0.9 \pm 0.5$~pb~\cite{d0ljxs},
where the dominant systematic uncertainties are on the $b$-tag efficiency and
the background estimate.
With 320~pb$^{-1}$, CDF measures $\sigma_{t\bar{t}} = 8.9\pm0.9\pm^{1.1}_{0.8}\pm0.5$~pb~\cite{cdfljxsbtag}.

\subsection*{All hadronic}

Even after satisfying a CDF multi-jet trigger requiring 
at least four jets with $E_{T}>$15~GeV and total transverse energy above 125~GeV, 
the expected signal-to-background ratio is only 1/3500.  
A kinematic selection exploits the higher transverse energy 
and more spherical distribution of jets from top quark decay to increase the S/B to 1/25.  
For events with between 6 and 8 jets in 311~pb$^{-1}$, 
CDF observes 816 $b$-tagged jets with an estimated background of 683$\pm$38 $b$-tagged jets.
The kinematic selection efficiency times branching ratio for $t\bar{t}$\ is about 7\%,
and on average there are 0.84 $b$-tags per $t\bar{t}$\ event.
The background $b$-tag rate is parameterized
from data with exactly 4 jets before the kinematic selection. 
With the excess ascribed to $t\bar{t}$\ production, CDF measures 
$\sigma_{t\bar{t}} = 8.0 \pm 1.7 \pm^{3.3}_{2.2} \pm 0.5$~pb~\cite{cdfallxs}.  
The dominant systematic uncertainty is the dependence of the selection efficiency
on the jet energy scale.
With 350~pb$^{-1}$, D0 measures $\sigma_{t\bar{t}} = 5.2\pm2.6\pm^{1.5}_{1.0}\pm0.3$~pb~\cite{d0allxs}.

\section*{Is there something new in Top Quark Decay?}

The standard model predicts $t \to W^{+}b$\ with 99.8\% branching fraction.  CDF and
D0 have performed several tests of non-standard model hypotheses.

{\bf Is there always a $b$ quark?}
From the relative rates and background estimate for 0, 1, and 2 $b$-tags in the lepton+jets and dilepton samples, 
one can extract the product of the 
$b$-tag efficiency and the branching ratio 
$R=\mathcal{BR}(t \to Wb)/\mathcal{BR}(t \to Wq)$, where $q$ is $d$, $s$, or $b$.  
With an independent estimate of the $b$-tag efficiency and 161~pb$^{-1}$, 
CDF sets a 95\% C.L.~limit that $R>$0.61~\cite{cdf_notvtb}. 
With 230~pb$^{-1}$, D0 sets a 95\% C.L.~limit that $R>0.64$~\cite{d0_notvtb}.

{\bf Is there always a $W^{+}$ boson?}    In the minimal supersymmetric standard model,
 the branching fraction for $t \to H^{+}b$\ is significant (above 10\%) for 
small and large values of $\tan \beta$. 
The $H^{+}$\ decays differently than a $W^{+}$ boson.  In particular,
$H^{+} \to \tau^{+}\nu_{\tau}$\ is enhanced at high $\tan \beta$, while
$H^{+} \to t^{\star}\bar{b} \to W^{+}b\bar{b}$\ is enhanced at low $\tan \beta$\ for 
a large Higgs mass.  CDF sets 95\% C.L.~limits in the plane of $\tan \beta$\ 
and $H^{+}$\ mass by comparison of the number of observed events in 
four final states (dilepton, dilepton with hadronic $\tau$\ decay, 
lepton+jets with one $b$-tag, lepton+jets with two $b$-tags) with
expectation from the minimal supersymmetric standard model~\cite{cdf_chargedhiggs}.

{\bf Is it $t \to W^{+}b$\ or $t \to W^{-}b$?}  If the top quark electric charge
is $-\frac{4}{3}$\ instead of $+\frac{2}{3}$, 
then the charge of the $W$ boson from top quark decay would be reversed.  D0 tests this hypothesis 
in 21 lepton+jets events with at least 4 jets and 2 $b$-tags in 365~pb$^{-1}$. 
This is a very pure sample with an estimated background of only 5\%. The lepton and $b$-jet combination is chosen with an estimated 79\% efficiency via a kinematic fit as the pairing most
consistent with a top quark mass of 175~GeV/$c^{2}$.  The magnitude
of the top quark electric charge is estimated as sum of the lepton charge
and the $b$-jet charge.  D0 models jet charge from $b\bar{b}$\ data. The information from the 
other top quark is not neglected, instead a second estimate of the top quark electric charge 
is obtained as the magnitude of the negative of the lepton charge and the second $b$-jet charge.  
Using a likelihood ratio test, D0 excludes the top quark electric charge $-\frac{4}{3}$\ 
hypothesis at 94\% C.L.~\cite{d0_topcharge}.

{\bf Is the $W^{+}$ helicity ``right''?}  70\% of the $W^{+}$ bosons from top quark decay
are expected to have a helicity (the projection of particle's spin onto its momentum vector) 
of zero due to the large top quark mass. The standard model, a $V\!-\!A$ theory, predicts that the other 
30\% are left-handed with helicity of $-1$.  However, if the top quark couples to new particles in the 
$t\!-\!W\!-\!b$\ vertex, then some fraction of the other 30\% could be right-handed with helicity $+1$.  
The emission angle of the charged lepton from the $W$ boson decay with respect to the direction of the $W$ 
boson in the top quark rest frame, $\cos \theta^{\star}$,\ is directly related to the helicity of the $W$ boson.  
Left-handed $W^{+}$ bosons preferentially emit the lepton in the opposite direction to the $W$ boost, 
and vice versa for right-handed $W^{+}$ bosons.  CDF and D0 have analysed distributions of lepton $p_{T}$\ 
and estimators for the reconstructed $\cos\theta^{\star}$ in the lepton+jets and dilepton data samples.  
All results are consistent within large statistical uncertainties with the standard model prediction~\cite{d0whel,cdfwhel}.

\section*{Does something new produce Top Quark Pairs?}

The tests in the previous section find that top quark decay is consistent 
with the standard model expectation.  With this assurance that measurements in different final states are indeed related by the standard model branching fractions, CDF and D0 combine
several measurements to obtain a more precise estimate of the top quark pair production cross section.  CDF finds $\sigma_{t\bar{t}} = 7.1 \pm 0.6 \pm 0.7 \pm 0.4$~pb~\cite{cdf_comboxs}, 
and D0 finds $\sigma_{t\bar{t}} = 7.1 \pm 1.2 \pm^{1.4}_{1.1} \pm 0.5$~pb~\cite{d0_comboxs} 
for a top quark mass of 175~GeV/$c^{2}$.  These measurements are in good agreement
with the prediction from NLO QCD.  

Having checked the total production rate,  CDF and D0 also search 
for indications of a resonance from a new massive particle, $X^{0}$, decaying into $t\bar{t}$.  
CDF and D0 reconstruct the invariant mass of the $t\bar{t}$\ system 
in the lepton+jets channel.  
CDF uses all events with 4 or more jets, 
D0 requires in addition at least one $b$-tag.  
In 370~pb$^{-1}$, D0 sees a slight excess below $m_{t\bar{t}}$\ = 450~GeV/$c^{2}$~\cite{d0_mttbar}. 
At the time of the conference with 320~pb$^{-1}$, CDF reported a  
2-standard deviation excess around $m_{t\bar{t}}$\ = 500~GeV/$c^{2}$.  
In a recent update with double the previous statistics, 
CDF now sees no excess in 680~pb$^{-1}$~\cite{cdf_mttbar}.
The 95\% C.L.~limit for $\sigma_{X^{0}} \times \mathcal{BR}(X^{0} \to t\bar{t})$\ is about 2~pb for $m_{t\bar{t}}$\ = 500~GeV/$c^{2}$, and about 0.5~pb for $m_{t\bar{t}}$\ = 800~GeV/$c^{2}$.

\section*{Does something new produce Single Top Quarks?}

Top quarks are also produced singly via the 
electroweak interaction. In the standard model, 
the single top quark production cross section,
0.88$\pm$0.11~pb in the s-channel and 1.98$\pm$0.25~pb in the t-channel
from the theoretical prediction~\cite{singletop_st}, 
is directly proportional to the CKM element $|V_{tb}|^{2}$,
and is only 3 times smaller than the pair production cross section.  
The s-channel is sensitive to new resonances like a new
massive $W'$\ boson, while the t-channel is sensitive to changes in
the $t\!-\!W\!-\!b$\ vertex like flavor-changing-neutral-currents~\cite{singletop_tait}.  
Furthermore, single top quark production is itself a background to the search for
the standard model Higgs boson via $WH$ production~\cite{cdfwh,d0wh}.

However, the experimental signature from single top quark production
(isolated lepton, missing transverse energy, two or more jets, one [t-channel] or two [s-channel] $b$-tags) 
is swamped by background from $W$+jets and top quark pair production.
D0 has developed several advanced multivariate techniques 
to discriminate single top quark production from backgrounds~\cite{d0_singletop_adv}.  
These require an excellent modeling of the background composition as well as 
the kinematics of both signal and background. In fact, for discovery 
of single top quark production, reduction of the systematic uncertainty 
from background modeling will be crucial.  
With a null hypothesis in 370~pb$^{-1}$,  D0 
excludes at 95\% C.~L. an s-channel cross section above 5.0~pb and 
a t-channel cross section above 4.4~pb, with expected limits of 3.3~pb and 4.3~pb respectively.  
This is a factor of 2-3 away from the expected standard model
cross sections, but is in the range of enhancements from physics beyond the standard model.

\section*{Precision Measurement of the Top Quark Mass}

Having proven that the observed top quark is consistent with the predictions of the standard model,
CDF and D0 make a precision measurement of the top quark mass. 

\begin{figure}
\begin{minipage}{0.49\textwidth}
\includegraphics[width=.99\textwidth]{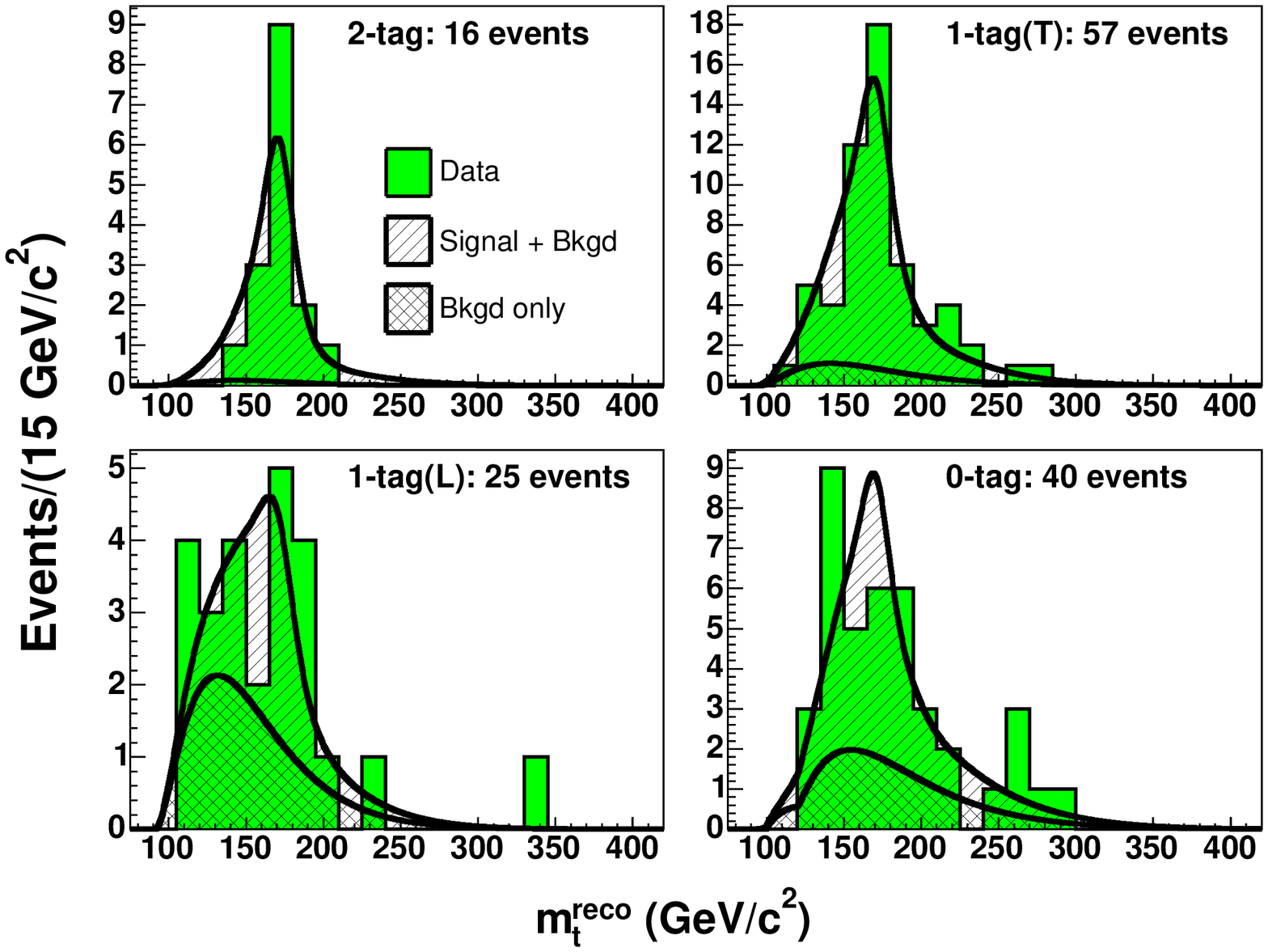}
\end{minipage}
\hspace{0.02\textwidth}
\begin{minipage}{0.49\textwidth}
\includegraphics[width=.99\textwidth]{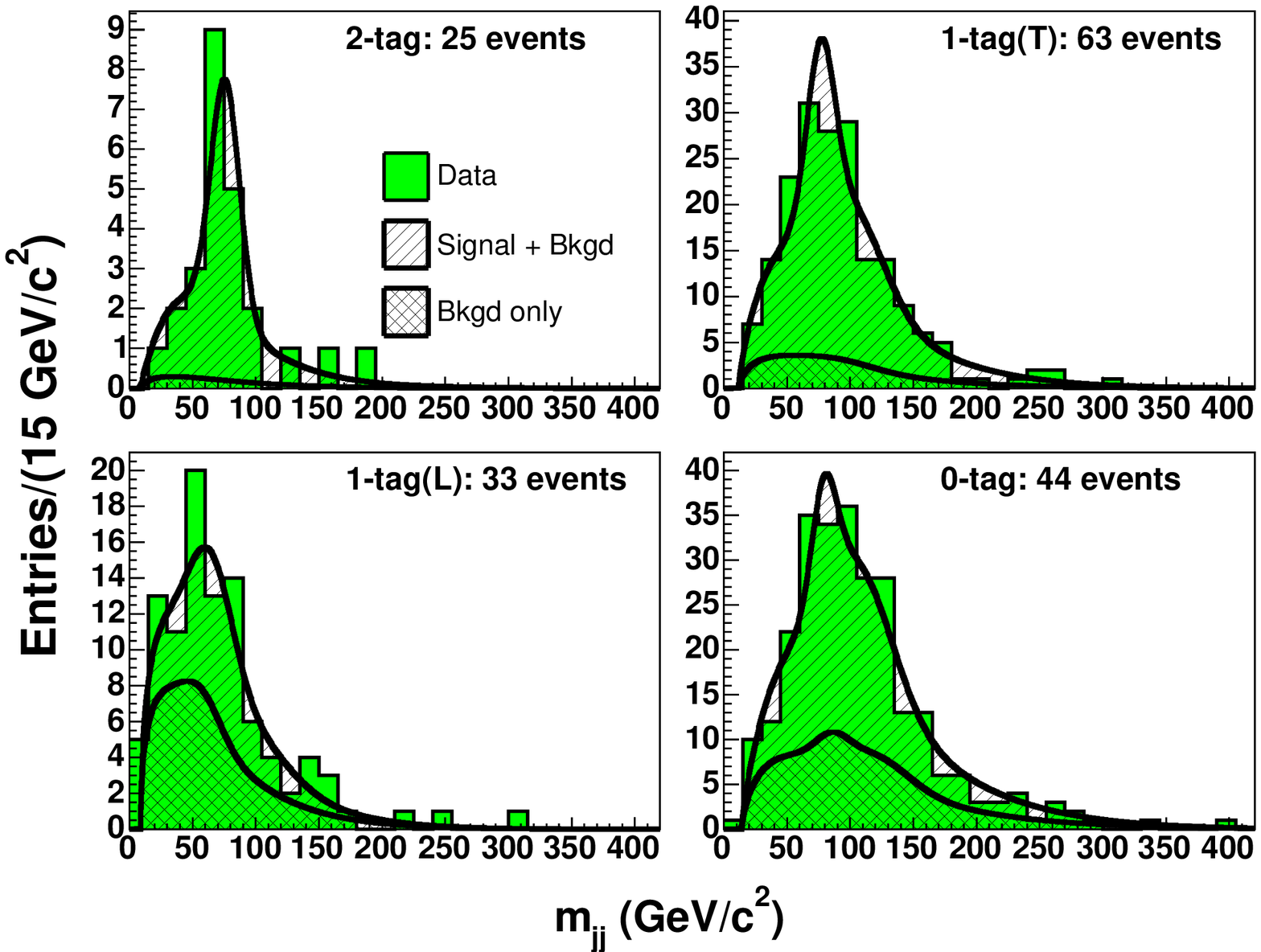}
\end{minipage}
  \caption{\label{fig:cdf_topmass} The reconstructed top quark mass (left) and
reconstructed W boson mass (right) distributions from 320~pb$^{-1}$\ 
CDF lepton+jets data compared with the best fit.}
\end{figure}

CDF performs a kinematic fit of lepton+jets events to the top quark pair production
and decay hypothesis in order to obtain improved resolution on the 
reconstructed top quark mass. For each event with 0/1/2 $b$-tags, there are 12/6/2 permutations
in the assignment of the four highest $E_{T}$\ jets to the partons from the top quark decay.
There are also two solutions for the neutrino $p_{z}$\ from the quadratic ambiguity in the $W \to \ell\nu$\ constraint.
The estimator for the top quark mass is the reconstructed top quark mass for the combination most consistent 
with the observed final state and the top quark pair production and decay hypothesis.
The top quark mass is extracted with a maximum likelihood 
fit of the observed reconstructed top quark
mass distribution to simulated distributions with various assumed values for the top quark mass.  
CDF divides a 320~pb$^{-1}$\ lepton+jets sample into 4 subsets depending on $b$-tag and jet multiplicity,
as shown in Figure~\ref{fig:cdf_topmass},
in order to optimize statistical sensitivity given 
the differences in signal-to-background and resolution. CDF measures a top quark mass of 
173.5$\pm^{2.7}_{2.6}\pm2.5 \pm 1.3$~GeV/$c^{2}$~\cite{cdf_topquarkmass}, 
where the first uncertainty is statistical, the second is from the jet energy scale, and
the third includes all other systematic uncertainties.

D0 uses the leading order matrix-element for $t\bar{t}$\ production and decay to calculate
the likelihood of the observed final state jets and lepton for each top quark mass hypothesis.
As the matrix-element requires the parton momenta, the probability for a parton energy to yield an
observed jet energy is parameterized from simulation.  
The current result does not use any $b$-tag information, 
instead the likelihoods from all 24 jet-parton and neutrino $p_{z}$\ assignments 
are multiplied together.  
In 320~pb$^{-1}$, D0 has 150 observed events with an estimated contribution
from $t\bar{t}$\ of 32$\pm$5\%.  D0 measures a top quark mass of 
169.5$\pm$3.0$\pm$3.2$\pm$1.7~GeV/$c^{2}$~\cite{d0_topquarkmass}.

CDF has developed the first application of a matrix-element technique to the dilepton channel.
The presence of two undetected neutrinos makes reconstructing the top quark mass particularly 
challenging.  With 33 events collected in 340 pb$^{-1}$, CDF measures a top quark mass of 
165.2$\pm$6.1$\pm$3.4 GeV/$c^{2}$~\cite{cdf_topquarkmass_dilepton}, 
the most precise single measurement of the top quark mass in the dilepton channel.  

The top quark mass measurement requires an excellent modeling of jet fragmentation and 
an excellent simulation of the calorimeter response to jets.
The approximate 3\% uncertainty on the CDF jet energy scale~\cite{cdf_jetenergy} 
translates into a 3 GeV/$c^{2}$\ uncertainty on the top quark mass.  
At low jet $E_{T}$, the dominant systematic on the 
jet energy scale is from the modeling of energy outside the jet cone for these broader jets, 
while at high $E_{T}$, the dominant systematic is from the calibration of the 
calorimeter response.  
The balance of a well-measured photon, or $Z \to \ell\ell$, against a recoiling jet provides
a cross check with limited statistics.

For the first time, with the higher statistics lepton+jets data samples, 
CDF and D0 are able to significantly constrain the jet energy scale from the 
invariant mass peak of the jets assigned to the $W \to q\bar{q'}$\ decay.
This observable is sensitive to the jet energy scale but does not
depend on the top quark mass. 
Figure~\ref{fig:cdf_topmass} also shows the CDF observed di-jet mass distribution 
from all permutations not including the $b$-tagged
jets.  CDF extracts a correction of -0.10$\pm^{0.78}_{0.80}$\ 
times the jet energy scale uncertainty, which is dependent on the jet $p_{T}$ and $\eta$.  
This is a correction of approximately -0.3\% to the jet energy scale, and reduces 
the systematic uncertainty on the top quark mass from the jet energy scale by 20\%.
In the matrix-element technique of D0, the parton energies are divided by a single correction factor,
which is determined by the constraint on the $W$ boson mass in the $t\bar{t}$\ matrix-element.
After calibration for limitations from the assumptions in the matrix-element technique,
D0 extracts a correction of 1.034$\pm$0.034.  This is a correction of approximately 3.4\% 
to the jet energy scale.  
All of the above calibrations are for non-$b$ jets.  
Both CDF and D0 estimate a systematic of approximately 0.6~GeV/$c^{2}$\ that accounts for relative differences 
between $b$-jets and light jets due to fragmentation models, 
semileptonic decay branching fractions, and color flow.   With higher statistics, $\gamma b$
and $Zb$ production, and even $Z \to b\bar{b}$\ for which CDF has a secondary vertex trigger,
may provide useful constraints.  

The Summer 2005 combination of the best measurements from CDF and D0 in each channel 
from Run-I and Run-II yields a world average value for the top quark mass of 
172.7$\pm$2.9~GeV/$c^{2}$~\cite{tev_topmass}, a 40\% improvement from 2004.  Via quantum loops, 
the $W$ boson mass is sensitive to the square of the top quark mass, the logarithm of 
the Higgs boson mass, and to new massive particles from beyond the standard model~\cite{weiglin}.  
With the full two-loop electroweak corrections~\cite{freitas}, 
the experimental uncertainty on the top quark mass is by a factor of two 
the dominant uncertainty on the standard model prediction for the $W$ boson mass:
the current 2.9~GeV/$c^{2}$\ uncertainty on the top quark mass corresponds to about 18~MeV/$c^{2}$\
on the $W$ boson mass prediction.  
In combination with other precision electroweak measurements, 
the improved measurement of the top quark mass sharpens the constraint on the 
undiscovered standard model Higgs boson mass to 91$\pm^{45}_{32}$~GeV/$c^{2}$~\cite{lepewk}. 
Including the direct search limit of 114.4~GeV/$c^{2}$\ from LEP~\cite{lephiggs}, 
the standard model Higgs boson mass is less than 219~GeV/$c^{2}$\ at 95\% C.L.

The future of the top quark mass measurement at the Tevatron
is bright, as the dominant systematic uncertainty from the jet energy scale now scales with 
statistics thanks to the \emph{in situ} constraint from $W \to q\bar{q'}$.  
With an integrated luminosity of 2~fb$^{-1}$\ per experiment and a conservative
assumption of no reduction in other systematic uncertainties, 
the projected top quark mass uncertainty is 1.7~GeV/$c^{2}$.
This is another 40\% reduction in uncertainty, and it will sharpen the constraint on the Higgs boson mass
even more at a very interesting time in particle physics, 
as the LHC turns on to search for the Higgs boson.
With 4~fb$^{-1}$\ and beyond, it is likely that the Tevatron precision measurement 
of the top quark mass will be comparable to the precision expected from the LHC~\cite{atlas_top}.  


\section*{Conclusion}

Thanks to the excellent performance
of the Tevatron accelerator complex and the CDF and D0 experiments,
the future of top quark physics at the Tevatron is bright.
The precision of the top quark mass measurement, 172.7$\pm$2.9~GeV/$c^{2}$\ 
has improved by 40\% in the last year alone.    
Although the observed top quark is consistent with the standard model top quark so far,
there is still lots of potential for surprises in the order of magnitude larger data samples 
currently under accumulation. Watch out for top results!


\section*{Acknowledgements}
I would like to thank the organizers of PANIC 2005 
and my colleagues on CDF and D0.  This work has been supported by U.~S.~Department of Energy contract DE-FG02-95ER40893.


\begin{thebibliography}{9}


\bibitem{pdgtopquark}
S.~Eidelman \emph{et al.}, 
\emph{Phys.~Lett.~B} \textbf{592}, 1 (2004). 
2005 partial update at \url{http://pdg.lbl.gov/}.

\bibitem{jacoreview}
D.~Chakraborty, J.~Konigsberg and D.~L.~Rainwater, 
\emph{Ann.~Rev.~Nucl.~Part.Sci.} \textbf{53}, 301 (2003). 

\bibitem{cdfdiscovery}
F.~Abe \emph{et al.}, The CDF Collaboration, \emph{Phys.~Rev.~Lett.} \textbf{74}, 2626 (1995). 

\bibitem{d0discovery} 
S.~Abachi \emph{et al.}, The D0 Collaboration, \emph{Phys.~Rev.~Lett.} \textbf{74}, 2632 (1995).

\bibitem{CDFD0TOP} Public web-pages and archive of conference notes 
from the Top Quark Physics Groups of 
CDF \url{http://www-cdf.fnal.gov/physics/new/top/top.html}
and D0
\url{http://www-d0.fnal.gov/Run2Physics/top/top_public_web_pages/top_public.html}.


\bibitem{ttbarMangano}
M.~Cacciari \emph{et al.}, JHEP \textbf{404}:68 (2004).

\bibitem{ttbarKidonakis} 
N.~Kidonakis and R.~Vogt, \emph{Phys.~Rev.~D} \textbf{68}, 114014 (2003).

\bibitem{electroncooling}
``Fermilab's Recycler beams take electron cooling to new heights'' in \emph{CERN Courier}, 
\textbf{45}, 7 (September 2005).

\bibitem{d0dileptonxs} D0 conference note 4850.
V.~M.~Abazov \emph{et al.}, The D0 Collaboration, \emph{Phys.~Lett.~B} \textbf{626}, 55 (2005).

\bibitem{cdfdileptonxs} CDF conference note 7942.
D.~Acosta \emph{et al.}, The CDF Collaboration, \emph{Phys.~Rev.~Lett.} \textbf{93}, 142001 (2004).

\bibitem{cdfelltau}
A.~Abulencia \emph{et al.}, The CDF Collaboration, submitted to \emph{Phys.~Rev.~Lett.}, hep-ex/0510063.

\bibitem{cdfljxs} CDF conference note 7753.
D.~Acosta \emph{et al.}, The CDF Collaboration, \emph{Phys.~Rev.~D} \textbf{72}, 052003 (2005).

\bibitem{d0ljxstopo} 
V.~M.~Abazov \emph{et al.}, The D0 Collaboration, \emph{Phys.~Lett.~B} \textbf{626}, 45 (2005).

\bibitem{d0ljxs} D0 conference note 4888.
V.~M.~Abazov \emph{et al.}, The D0 Collaboration, \emph{Phys.~Lett.~B} \textbf{626}, 35 (2005).

\bibitem{cdfljxsbtag} CDF conference note 7801.
D.~Acosta \emph{et al.}, The CDF Collaboration, \emph{Phys.~Rev.~D} \textbf{71}, 052003 (2005).

\bibitem{cdfallxs} CDF conference note 7793.

\bibitem{d0allxs} D0 conference note 4879.

\bibitem{cdf_notvtb}
D.~Acosta \emph{et al.}, The CDF Collaboration, \emph{Phys.~Rev.~Lett.} \textbf{95}, 102003 (2005).

\bibitem{d0_notvtb} DO conference note 4833.

\bibitem{cdf_chargedhiggs}
A.~Abulencia \emph{et al.}, The CDF Collaboration, \emph{Phys.~Rev.~Lett.} \textbf{96}, 042003 (2006).

\bibitem{d0_topcharge} DO conference note 4876.

\bibitem{d0whel} D0 conference note 4839.
V.~M.~Abazov \emph{et al.}, The D0 Collaboration, \emph{Phys.~Rev.~D Rap.~Comm.} \textbf{72}, 011104(R) (2005).

\bibitem{cdfwhel}
A.~Abulencia \emph{et al.}, The CDF Collaboration, submitted to \emph{Phys.~Rev.~Lett.}, hep-ex/0511023.

\bibitem{cdf_comboxs} CDF conference note 7794.

\bibitem{d0_comboxs} D0 conference note 4906.

\bibitem{cdf_mttbar} CDF conference notes 7971, 8087.

\bibitem{d0_mttbar} D0 conference note 4880.

\bibitem{singletop_st}  
B.~W.~Harris \emph{et al.}, \emph{Phys.~Rev.~D} \textbf{66}, 054024 (2002). 

\bibitem{singletop_tait} 
T.~M.~P.~Tait and C.~-P.~Yuan, \emph{Phys.~Rev.~D} \textbf{63}, 014018 (2001). 

\bibitem{cdfwh}
A.~Abulencia \emph{et al.}, The CDF Collaboration, accepted by \emph{Phys.~Rev.~Lett.}, hep-ex/0512051.

\bibitem{d0wh} DO conference note 4896.
V.~M.~Abazov \emph{et al.}, The D0 Collaboration, \emph{Phys.~Rev.~Lett.} \textbf{94}, 091802 (2005).

\bibitem{d0_singletop_adv} D0 conference notes 4722, 4871.
V.~M.~Abazov \emph{et al.}, The D0 Collaboration, \emph{Phys.~Lett. B} \textbf{622}, 265-276 (2005).

\bibitem{cdf_topquarkmass}
A.~Abulencia \emph{et al.}, The CDF Collaboration, \emph{Phys.~Rev.~Lett.} \textbf{96}, 022004 (2006).\\
A.~Abulencia \emph{et al.}, The CDF Collaboration, submitted to \emph{Phys.~Rev.~D}, hep-ex/0510048.\\
A.~Abulencia \emph{et al.}, The CDF Collaboration, submitted to \emph{Phys.~Rev.~D}, hep-ex/0512009.

\bibitem{d0_topquarkmass} D0 conference note 4874.

\bibitem{cdf_topquarkmass_dilepton}
A.~Abulencia \emph{et al.}, The CDF Collaboration, submitted to \emph{Phys.~Rev.~Lett.}, hep-ex/0512070.

\bibitem{cdf_jetenergy}
A.~Bhatti \emph{et al.}, submitted to \emph{Nucl.~Instr.~Meth.}A, hep-ex/0510047.

\bibitem{tev_topmass} The CDF and D0 Collaborations, and the Tevatron Electroweak Working Group, hep-ex/0507091.

\bibitem{weiglin} S.~Heinemeyer and G.~Weiglin, \url{http://quark.phy.bnl.gov/~heinemey/uni/plots/}.

\bibitem{freitas} M.~Awramik \emph{et al.}, \emph{Phys.~Rev.} D \textbf{69}, 053006 (2004). 

\bibitem{lepewk} The ALELPH, DELPHI, L3, and OPAL Collaborations, and the LEP Electroweak Working Group, hep-ex/0511027.

\bibitem{lephiggs} The ALELPH, DELPHI, L3, and OPAL Collaborations, and the LEP Higgs Working Group, 
\emph{Phys.~Lett. B} \textbf{565}, 61-75 (2003).


\bibitem{atlas_top} I. Borjanovic \emph{et al.}, hep-ex/0403021. CERN Yellow Report 2000-004. 

\end{thebibliography}
\end{document}